\documentclass[twocolumn]{article}

\usepackage{graphics}
\usepackage{amssymb}

\author{A.~Sherman$^1$ and M.~Schreiber$^2$\\
$^1$Institute of Physics, University of Tartu, Riia 142,
 51014 Tartu, Estonia\\
$^2$Institut f\"ur Physik, Technische Universit\"at
Chemnitz, D-09107 Chemnitz, Germany}

\title{Magnetic susceptibility in the pseudogap phase of cuprate perovskites}

\begin{document}

\maketitle

{\bf Abstract} We calculate the magnetic susceptibility in the pseudogap phase of cuprates using the two-dimensional $t$-$J$ model and the Mori projection operator technique. In this phase, the Fermi arcs lead to a quasi-elastic incommensurate magnetic response for low temperatures. The theory accounts for the reorientation of the susceptibility maxima from the axial to the diagonal direction occurring at small hole concentrations in La$_{2-x}$Sr$_x$CuO$_4$. A small cusp of the hole dispersion near the Fermi level, which is connected with the spin-polaron band, affords the growth of the maxima with decreasing frequency. As in the superconducting phase, the susceptibility maxima have an hourglass dispersion. The assumption of the phase separation into regions of superconducting and pseudogap phases in underdoped cuprates explains the lack of the superconducting gap in their susceptibility. The observed rapid change of widths of the susceptibility maxima with frequency is related to the contribution of the pseudogap phase. This contribution may also explain the strengthening of the quasi-elastic response by modest magnetic fields and by small impurity concentrations.

\section{Introduction}
\label{intro}
The magnetic susceptibility of underdoped cuprates differs from that of more heavily doped crystals by the lack of the so-called superconducting gap -- the suppression of the imaginary part of the susceptibility $\chi''$ at low frequencies $\omega$ in the superconducting state \cite{Rossat,Kofu,Haug}. In underdoped samples, for temperatures below the superconducting transition temperature $T_c$ the susceptibility is finite up to the lowest measured frequencies, and in some experiments it demonstrates even the tendency for a growth with decreasing $\omega$ \cite{Kofu}. This behavior is reflected in the quasi-elastic magnetic neutron scattering observed in lanthanum and yttrium cuprates.

In the considered range of hole concentrations $x$ and temperatures $T$ the crystals are either in the superconducting or in the pseudogap phases. The superconducting gap in the susceptibility is well reproduced in different theoretical approaches (see, e.g., \cite{Lavagna,Brinckmann,Sherman03}). To our knowledge the mechanism of the quasi-elastic magnetic response was much less studied. In Ref.~\cite{Kofu}, based on the sharp difference in the spin correlation lengths below and above some small frequency it was supposed that below $T_c$ the underdoped crystal La$_{2-x}$Sr$_x$CuO$_4$ (LSCO) is phase-separated into superconducting and non-su\-per\-con\-duc\-ting regions. The latter regions are in the pseudogap phase and it can be inferred that they are responsible for the low-frequency magnetic response.

The mechanism of this response may be the following. In comparison with the superconducting phase the pseudogap state is characterized by an increased damping of hole states and by Fermi arcs -- fragments of the Fermi surface near the nodal points \cite{Norman,Yoshida}. For frequencies below the resonance frequency $\omega_r$ the locations of maxima of the magnetic susceptibility in the Brillouin zone are determined by peaks of the spin-excitation damping \cite{Sherman11}. The main contribution to these peaks is made by the process of spin-excitation transformation into a fermion pair. In the superconducting phase the respective states are located on equal energy contours of the fermion dispersion, the size of which decreases with decreasing $\omega$. This leads to a susceptibility, which vanishes below some frequency. In contrast, in the pseudogap phase due to the Fermi arcs the length of the equal energy contours remains finite up to zero frequency, that ensures a finite susceptibility up to $\omega\sim T$. Thus, indeed it can be expected that in the phase-separated state the low-frequency susceptibility is determined by regions of the pseudogap phase.

This picture permits to reach some other conclusions about the magnetic response of the pseudogap state. In the superconducting phase, for frequencies immediately below $\omega_r$ the number of fermion pairs, which contribute to the susceptibility on the edges of the Brillouin zone, is larger than that contributing on the diagonals \cite{Brinckmann,Sherman11}. As a consequence the maxima of $\chi''$ are located on the zone edges at the momenta $(\pi\pm\kappa,\pi)$, $(\pi,\pi\pm\kappa)$, as it is indeed observed in moderately doped cuprates \cite{Yoshizawa,Birgeneau} (here and hereafter we set the lattice period $a$ of a CuO$_2$ plane as the unit of length). The same situation occurs in the pseudogap phase, in which for moderate $x$ the susceptibility maxima are peaked at the zone edges down to $\omega=0$. However, in LSCO the Fermi arc length is approximately proportional to the hole concentration \cite{Yoshida}. As the arc length decreases, the number of fermion states contributing to the susceptibility on the zone edges is diminished faster than that contributing on the diagonals. As a result for small $x$ and $\omega$ the susceptibility maxima shift to the diagonals of the Brillouin zone, as it is indeed observed in the pseudogap phase of LSCO for $x\lesssim 0.05$ \cite{Fujita11}. For these small hole concentrations as well as for larger $x$ the dispersion of the maxima has an hourglass shape, in agreement with experimental observations \cite{Fujita11}.

If the hole dispersion has no peculiarities at the Fermi level, the susceptibility $\chi''$ in the pseudogap phase decreases monotonously with frequency, remaining finite up to $\omega\sim T$. In contrast, in Ref.~\cite{Kofu} a growth of the susceptibility of an underdoped LSCO was observed as the frequency decreases. In the considered theory such a behavior can be obtained if there is a cusp in the hole dispersion at the Fermi level. Such a cusp was found in calculations carried out in the two-dimensional (2D) $t$-$J$ model at moderate $x$ \cite{Sherman97}. In this case the cusp is connected with the spin-polaron band. A similar cusp is supposedly observed in some photoemission spectra of underdoped cuprates (see, e.g., Fig.~1 in Ref.~\cite{Terashima}).

In the pseudogap phase, the Fermi arcs produce a peculiar shape of the susceptibility $\chi''$, which as a function of wave vector has a maximum and an extended shoulder. The shoulder is weak for small frequencies. However, it grows rapidly with $\omega$ and determines the width of the spectrum for $\omega>5$~meV. As a consequence the width changes steeply near this frequency, in close analogy with experimental observations in LSCO \cite{Kofu}.

The structure of the article is the following. The main formulas used in the calculations are given in the next section. Obtained results are considered in Sec.~\ref{results}. Section~\ref{conclusion} is devoted to the discussion of possible applications of the obtained results for the interpretation of the strengthening of the quasi-elastic response by the magnetic field and impurities, and to the concluding remarks.

\section{Main formulas}
\label{formulas}
General formulas for the magnetic susceptibility of the 2D $t$-$J$ model in the superconducting phase were derived in Ref.~\cite{Sherman11} with the use of the Mori projection operator formalism \cite{Mori}. The reformulation for the pseudogap phase needs only minor changes. Therefore, we adduce merely equations, which were used in the calculations.

In the Mori formalism the susceptibility is described by the equation
\begin{equation}\label{chi}
\chi({\bf k}\omega)=-\frac{h_{\bf k}}{\omega^2-\omega\Pi({\bf k}\omega)-\omega_{\bf k}^2}.
\end{equation}
In this equation,
\begin{equation}\label{hk}
h_{\bf k}=4\left(tF_1+JC_1\right)\left(\gamma_{\bf k}-1\right),
\end{equation}
and
\begin{equation}\label{wk}
\omega^2_{\bf k}=16J^2\alpha|C_1|\left(1-\frac{tF_1}{J\alpha|C_1|} \right)\left(1-\gamma_{\bf k}\right)\left(\delta+1+\gamma_{\bf k}\right),
\end{equation}
where {\bf k} is the 2D wave vector, $t$ and $J$ are the hopping and exchange constants between neighboring sites, $\gamma_{\bf k}=\left[ \cos(k_x)+\cos(k_y)\right]/2$, $$F_1=\frac{1}{N}\sum_{\bf k}\gamma_{\bf k}\left\langle a^\dagger_{\bf k\sigma}a_{\bf k\sigma}\right\rangle\;{\rm and}\; C_1=\frac{1}{N}\sum_{\bf k}\gamma_{\bf k}\left\langle s^{+}_{\bf k}s^{-}_{\bf k}\right\rangle$$
are correlators of the hole $a^{(\dagger)}_{\bf k\sigma}$ and spin-$\frac{1}{2}$ $s^\pm_{\bf k}$ operators on the neighboring sites, $\sigma=\pm 1$ is the spin projection, the angular brackets denote the statistical averaging and $N$ is the number of sites. The parameter $\alpha$ serves for correcting the decoupling procedure \cite{Kondo,Shimahara}. For small $x$ it is approximately equal to $1.7$ \cite{Sherman03a}. The parameter $\delta$ describes the gap in the spin-excitation spectrum near the antiferromagnetic wave vector ${\bf Q}=(\pi,\pi)$ due to the finite temperature \cite{Shimahara} and/or the hole concentration \cite{Sherman03a}. The magnitude of this gap is directly connected with the correlation length of the short-range antiferromagnetic order and it grows with $x$ and $T$. In some cases the magnitude of the gap determines the frequency $\omega_r$ of the waist in the hourglass dispersion of the susceptibility maxima \cite{Sherman11}.

There are four contributions to the polarization operator $\Pi({\bf k}\omega)$ of spin excitations in Eq.~(\ref{chi}). These contributions are connected with the processes of the spin-excitation decay into an electron-hole pair, the decays into a pair assisted by a hole or by another spin excitation, and the decay into three spin excitations. For small frequencies considered in this article the first of the mentioned processes dominates. Its contribution to the imaginary part of the polarization operator reads
\begin{eqnarray}
&&{\rm Im}\Pi({\bf k}\omega)=\frac{2\pi}{N\omega h_{\bf k}}\sum_{\bf k'}f^2({\bf kk'})\bigg[ N_F(\varepsilon_{\bf k+k'},\varepsilon_{\bf k'})Z^2\nonumber\\
&&\quad\times\delta(\omega+\varepsilon_{\bf k+k'}-\varepsilon_{\bf k'}) +\bigg(\frac{ZZ'}{8t-2\Delta}\bigg)^2U(\omega)\nonumber\\
&&\quad+N_F(\varepsilon_{\bf k+k'},\varepsilon_{\bf k+k'}+\omega)\frac{ZZ'}{8t-2\Delta} S(\varepsilon_{\bf k+k'}+\omega) \nonumber\\
&&\quad+N_F(\varepsilon_{\bf k'}-\omega,\varepsilon_{\bf k'})\frac{ZZ'}{8t-2\Delta} S(\varepsilon_{\bf k'}-\omega)\bigg], \label{impi}
\end{eqnarray}
where
\begin{eqnarray}
f({\bf kk'})&=&\frac{1+x}{2}\bigg[\varphi({\bf -k'-k,k})-\frac{1}{2}\varphi({\bf  -k'-k},{\bf 0})\nonumber\\
&&\quad\quad\quad-\frac{1}{2}\varphi({\bf k'},{\bf 0})\bigg],\label{f}\\
\varphi({\bf kk'})&=&t_{\bf k}t_{\bf k+k'}-4t^2\gamma_{\bf k'}-4(t')^2\gamma'_{\bf k'},\label{phi}
\end{eqnarray}
$t'$ is the hopping constant between next-nearest neighbor sites, $\gamma'_{\bf k}=\cos(k_x)\cos(k_y)$, $t_{\bf k}=4t\gamma_{\bf k}+4t'\gamma'_{\bf k}$, $N_F(\omega,\omega')=n_F(\omega)-n_F(\omega')$, $n_F(\omega)= \left(e^{\omega/T}+1\right)^{-1}$. Equation~(\ref{impi}) was obtained in the supposition that the hole spectral function is described by the expression
\begin{equation}\label{A}
A({\bf k}\omega)=Z\delta(\omega-\varepsilon_{\bf k})+\frac{Z'}{8t-2\Delta}S(\omega),
\end{equation}
where
\begin{equation}\label{S}
S(\omega)=\left\{\begin{array}{ll}
            1 & \mbox{if $\varepsilon_{\bf  Q}-2t<\omega<-\Delta$,}\\
              & \mbox{or $\Delta<\omega<\varepsilon_{\bf  Q}+6t$,}\\
            0 & \mbox{in other cases,}
\end{array}\right.
\end{equation}
$\varepsilon_{\bf  k}$ is the hole dispersion and $\Delta$ is the pseudogap magnitude. The first term on the right-hand side of the spectral function (\ref{A}) corresponds to the coherent peak and the second term to the incoherent continuum of the hole-spin-excitation scattering. Such a form of the $A({\bf k}\omega)$ is typical for the $t$-$J$ model (see, e.g., \cite{Sherman03a}). The continuum spans the frequency range approximately equal to the width of the initial band $8t$. For moderate doping the lower edge of the continuum lies approximately $2t$ below the bottom of the hole band $\varepsilon_{\bf  k}$, which is located at the momentum ${\bf  Q}$ in the used hole picture. The spectral weights of the coherent and incoherent parts of the spectral function are connected by the relation \cite{Sherman11}
$$Z+Z'=\frac{1+x}{2}.$$
\begin{figure}
\centerline{\resizebox{0.75\columnwidth}{!}{\includegraphics{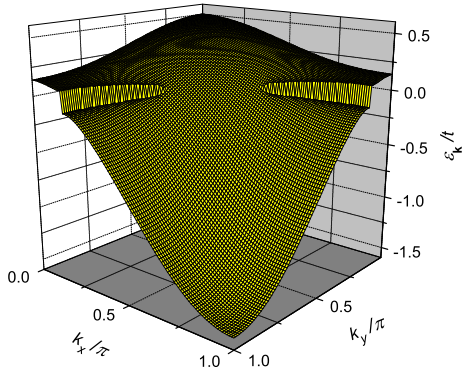}}}
\caption{The hole dispersion in the pseudogap phase, Eq.~(\protect\ref{dispersion}), in the first quadrant of the Brillouin zone for $l=1$ and $\Delta=0.1t$. Vertical fragments of the dispersion near the Fermi level correspond to gapped regions.} \label{Fig1}
\end{figure}

To simplify further calculations we set $T=0$ (we shall discuss possible consequences of a nonzero temperature in the next section). For $T=0$ and small positive frequencies the quantity
$$U(\omega)= \int_{-\infty}^{\infty}N_F(\omega',\omega+\omega')S(\omega') S(\omega+\omega') d\omega'$$
in Eq.~(\ref{impi}) reads
$$U(\omega)=\left\{\begin{array}{cl}
            0 & \mbox{if $\omega<2\Delta$,}\\
            \omega-2\Delta & \mbox{if $2\Delta<\omega<2t-\varepsilon_{\bf  Q}+\Delta$}
\end{array}\right.$$
The real part of $\Pi({\bf k}\omega)$ is calculated using the Kramers-Kronig relation.

To imitate the hole dispersion in the pseudogap phase we use the band
\begin{equation}
\frac{E_{\bf  k}}{t}=\frac{1}{2}\Big[\cos(k_x)+\cos(k_y)\Big]
-0.3\cos(k_x)\cos(k_y)-0.2,\label{band}
\end{equation}
which has the Fermi surface similar to that observed in La$_{1.88}$Sr$_{0.12}$CuO$_4$ \cite{Yoshida}, and construct $\varepsilon_{\bf k}$ as
\begin{equation}\label{dispersion}
\varepsilon_{\bf k}=\left\{\begin{array}{cl}
            \sqrt{E^2_{\bf k}+\Delta^2_{\bf k}}& \mbox{if $E_{\bf k}\geq 0$,}\\[0.7ex]
            -\sqrt{E^2_{\bf k}+\Delta^2_{\bf k}}& \mbox{if $E_{\bf k}< 0$.}
\end{array}\right.
\end{equation}
In Eq.~(\ref{dispersion}), $\Delta_{\bf k}$ has the shape of the $d$-wave gap function, $\widetilde{\Delta}_{\bf k}=\Delta[\cos(k_x)-\cos(k_y)]/2$, everywhere except circular regions around the nodal points ${\bf q}=(\pm\kappa,\pm\kappa)$, $E_{\bf q}=0$. In these regions $\Delta_{\bf k}=0$. Apparently the diameter $l$ of these regions is close to the arc length. The hole dispersion $\varepsilon_{\bf k}$ obtained in such a way is shown in Fig.~\ref{Fig1}.

\section{Results and discussion}
\label{results}
\begin{figure}
\centerline{\resizebox{0.75\columnwidth}{!}{\includegraphics{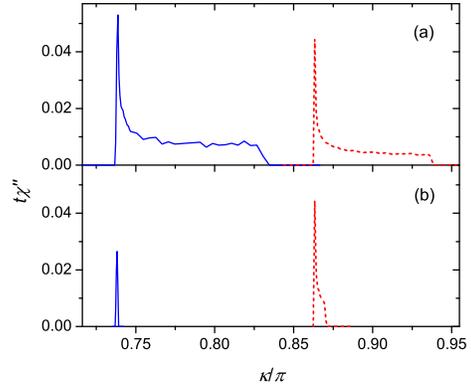}}}
\caption{Momentum dependencies of $\chi''({\bf k}\omega)$ for $\omega=0.001t$ along the edge [blue solid lines, ${\bf k}=(\pi,\kappa)$] and along the diagonal [red dashed lines, ${\bf k}=(\kappa,\kappa)$] of the Brillouin zone. The parameter $l=2$ in panel (a) and 0.6 in panel (b). Other parameters are given in the text.} \label{Fig2}
\end{figure}

Typical momentum dependencies of the imaginary part of the susceptibility obtained with the above formulas for a small frequency are given in Fig.~\ref{Fig2}. For this figure the following parameters were used: $J=0.2t$, $\delta=0.25$, $\Delta=0.05t$ and $Z/Z'=0.1$.
The parameter $\delta$ was chosen to obtain the frequency of the hourglass waist $\omega_r\approx 40$~meV, which is close to the value observed in moderately doped cuprates \cite{Fujita11}. This frequency is near the magnitude of the gap in the spin-excitation spectrum at {\bf Q}, mentioned in the previous section, if $\omega_r$ falls into the region of small spin-excitation damping below the edge of the electron-hole continuum at {\bf Q} \cite{Sherman11}. Here and hereafter changing to energy units we proceed from the estimate $t=500$~meV \cite{McMahan,Gavrichkov}. Thus, $J=100$~meV and $\Delta=25$~meV, which are also close to the values in moderately doped cuprates \cite{Yoshida,Fujita11}. Ratios $Z/Z'\sim 0.1$ were obtained in the self-consistent calculations in the 2D $t$-$J$ model for hole concentrations $0.07\lesssim x\lesssim 0.15$ \cite{Sherman99}. In LSCO the Fermi arc length decreases nearly linearly with $x$ \cite{Yoshida}. Using the experimental data of Ref.~\cite{Yoshida} we can relate the value $l=2$ to $x\approx 0.12$ and $l=0.6$ to $x\approx 0.04$. For the latter hole concentration the parameter $\delta$ has to be decreased, since the value of the resonance frequency decreases with $x$ \cite{Sherman03,Fujita11}, while the pseudogap magnitude $\Delta$ has to be increased \cite{Yoshida,Sherman99}. However, for small frequencies these changes lead only to minor modifications in the dependencies in Fig.~\ref{Fig2}.

In the calculations, we have approximated the $\delta$-fun\-c\-ti\-on in the first term of Eq.~(\ref{impi}) by the step function
$$\bar{\delta}(\omega,\eta)=\left\{\begin{array}{cl}
            (2\eta)^{-1} & \mbox{if $-\eta<\omega<\eta$,}\\
            0 & \mbox{in other cases.}
\end{array}\right.
$$
It is one of the representations of the $\delta$-function, which describes the convolution of the coherent peaks in $A({\bf k}\omega)$ better than the usually used Lorentzian representation, if the peaks have no extended tails. As in the superconducting phase, if the Lorentzian representation is used, the incommensurate magnetic response is obtained only for a pronounced nesting of the Fermi surface. The $\Pi$-shaped representation allows one to get such a response without the nesting. This result is in conformity with experimental observations of the incommensurate response in crystals, in which Fermi surfaces derived from photoemission have no obvious nesting (see Ref.~\cite{Sherman11} and references therein). As follows from Eq.~(\ref{S}), we set the intensity of the incoherent continuum equal to zero in the range $-\Delta < \omega < \Delta$. An increase of this intensity leads to a growth of $\chi''({\bf Q})$, which finally suppresses the incommensurate peaks when the intensity becomes comparable to that outside the range. In the calculations, we set $\eta=\omega/3$ for $\omega<0.005t$ and $\eta=0.002t$ for larger frequencies. An increase of the width $\eta$ as well as the Fermi arc length results in the growth of $\chi''({\bf Q})$.

As seen in Fig.~\ref{Fig2}, the low-frequency magnetic response is incommensurate -- the maxima on the symmetry lines of the Brillouin zone are displaced from {\bf Q}. For the considered small frequency this incommensurability is connected with maxima in the spin-excitation damping (\ref{impi}) descended from the first term on its right-hand side. Indeed, from Eq.~(\ref{chi}) we get
\begin{equation}\label{imchi}
\chi''({\bf k}\omega)=-\frac{\omega h_{\bf k}{\rm Im}\Pi({\bf k}\omega)}{\left[\omega^2 -\omega{\rm Re}\Pi({\bf k}\omega)-\omega^2_{\bf k}\right]^2+\left[\omega{\rm Im}\Pi({\bf k}\omega)\right]^2}.
\end{equation}
The susceptibility above the resonance frequency is mainly determined by the denominator of this equation, and below $\omega_r$ by its numerator. For small $\omega$
$$\chi''({\bf k}\omega)\approx -\omega h_{\bf k}\omega^{-4}_{\bf k}{\rm Im}\Pi({\bf k}\omega),$$
where $h_{\bf k}$, Eq.~(\ref{hk}), is a slowly varying function near {\bf Q}, while $\omega^{-4}_{\bf k}$ is strongly peaked at this wave vector [see equation~(\ref{wk})]. Thus, the incommensurate peaks may be related only to the last multiplier in this equation. The fermion states with the energies $\varepsilon_{\bf k}=\pm\omega/2$ make the main contribution to the peaks in ${\rm Im}\Pi({\bf k}\omega)$.
\begin{figure}
\centerline{\resizebox{0.65\columnwidth}{!}{\includegraphics{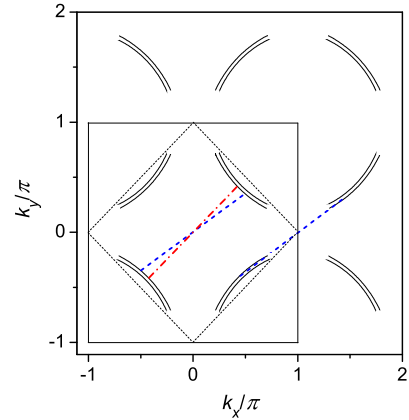}}}
\caption{The contour plot of the dispersion $\varepsilon_{\bf k}$ for $l=2$ and $\Delta=0.05t$. Contours correspond to the energies $\pm 0.03t$. The squares shown by thin solid and dotted lines are the first and magnetic Brillouin zones. The blue dashed and red dash-dotted lines connect electron-hole pairs contributing to the susceptibility on the edge and on the diagonal of the Brillouin zone, respectively.} \label{Fig3}
\end{figure}
The location of these states in the momentum space is shown in Fig.~\ref{Fig3}. Here arcs are sections of the hole dispersion $\varepsilon_{\bf k}$ at the energies $\pm 0.03t$. The blue dashed and red dash-dotted lines connect electron-hole pairs, which make the main contribution to the magnetic susceptibility on the edge and on the diagonal of the Brillouin zone, respectively. The location of these states is restricted to the vicinity of the arcs, and the difference in the wave vectors of the states is equal to the spin-excitation momentum. As a consequence the low-frequency susceptibility is sharply peaked at certain regions of the Brillouin zone, as seen in Fig.~\ref{Fig2}. In Fig.~\ref{Fig3}, opposite sides of the magnetic Brillouin zone are displaced by the vector {\bf Q} from each other. As seen from the figure, the line segments connecting the electron-hole pairs differ from {\bf Q} and, therefore, the response is incommensurate. Notice that there are two groups of decay processes for the susceptibility on the edge of the Brillouin zone (the two dashed lines in Fig.~\ref{Fig3}) and only one group for the diagonal (the dash-dotted line). This fact explains the larger value of $\chi''$ on the edge in comparison with the diagonal for the parameters of Fig.~\ref{Fig2}(a). As mentioned above, in moderately underdoped cuprates the maxima of the low-frequency magnetic susceptibility are also located on the zone edges \cite{Yoshizawa,Birgeneau,Fujita11}.

\begin{figure}
\centerline{\resizebox{0.75\columnwidth}{!}{\includegraphics{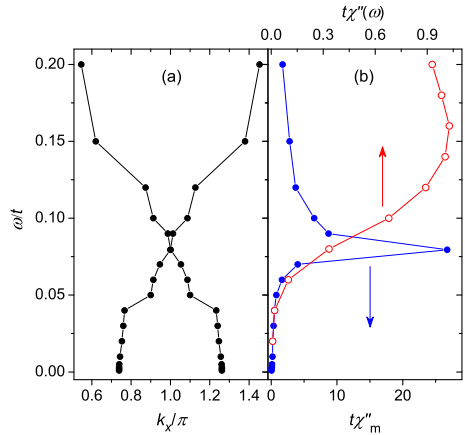}}}
\caption{(a) The dispersion of the maxima in $\chi''({\bf k}\omega)$ at the edge of the Brillouin zone, ${\bf k}=(k_x,\pi)$, for $l=2$ and $\Delta=0.05t$. Other parameters are given in the text. (b) The value of $\chi''({\bf k}\omega)$ at the maximum (filled blue circles) and the local susceptibility $\chi''(\omega)$ (open red circles) as functions of frequency.} \label{Fig4}
\end{figure}

With increasing $\omega$ the influence of the Fermi arcs on the susceptibility falls off and $\chi''$ approaches the value it would have in the superconducting phase for the same parameters. As a consequence the incommensurate maxima of the susceptibility have the dispersion of the hourglass shape, which is similar to that in the superconducting phase (see, e.g., Fig.~4 in Ref.~\cite{Sherman11}). This dispersion is shown in Fig.~\ref{Fig4}(a).
Notice that this shape of the dispersion is retained also for smaller lengths of the Fermi arcs, when the susceptibility at very small frequencies is peaked on the diagonals [see Fig.\ref{Fig2}(b)]. As in the superconducting phase, the downward-directed branch of the dispersion is connected with the numerator of Eq.~(\ref{imchi}), and the upward-directed branch with its denominator. For $\omega > \omega_r$ the momentum dependence of $\chi''$ is much more isotropic than below the resonance frequency. Since by definition (\ref{chi}) the susceptibility coincides with Green's function of spin excitations, the upward-directed branch displays the dispersion of these excitations. In Fig.~\ref{Fig4}(b) the frequency dependencies of $\chi''({\bf k}\omega)$ at the maximum and of the local susceptibility $$\chi''(\omega)=N^{-1}\sum_{\bf k}\chi''({\bf k}\omega)$$
are shown. As in the superconducting phase, the susceptibility maximum is most intensive at the waist frequency. The maximum of the local susceptibility is shifted to higher frequencies due to a larger susceptibility in the central part of the Brillouin zone in this frequency range.

\begin{figure}
\centerline{\resizebox{0.85\columnwidth}{!}{\includegraphics{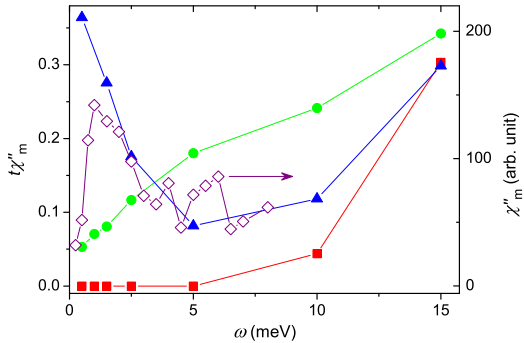}}}
\caption{The susceptibility $\chi''({\bf k}\omega)$ at the incommensurate maximum on the edge of the Brillouin zone as a function of frequency in the pseudogap (green circles, $l=2$) and superconducting (red squares) phases. Blue triangles show the susceptibility in the pseudogap phase in the case of the cusp in the hole dispersion at the Fermi level when $l=2$. In all cases $\Delta=0.05t$, other parameters are given in the text. The experimental susceptibility in La$_{1.875}$Sr$_{0.125}$CuO$_4$ measured at $T=4$~K \protect\cite{Kofu} is shown by purple rhombuses for comparison (the right axis).} \label{Fig5}
\end{figure}

The main difference of the dispersion in Fig.~\ref{Fig4} from that in the superconducting phase (see Fig.~4 in Ref.~\cite{Sherman11}) is the absence of the superconducting gap in the susceptibility at small frequencies. The appearance of this gap can be understood in the following manner. By analogy with the discussion of Fig.~\ref{Fig3}, in the superconducting phase the main contribution to the susceptibility at the frequency $\omega<\omega_r$ is made by fermion states with the energies $\xi_{\bf k}=(E_{\bf k}^2+\widetilde{\Delta}_{\bf k}^2)^{1/2} \approx\omega/2$. These states are located near the boundaries of crescent pockets, the shape of which resembles arcs in Fig.~\ref{Fig3}. However, in contrast to these arcs the length of the pockets shrinks with decreasing $\omega$. As a result intensities of the susceptibility maxima are diminished. Maxima on the edges of the Brillouin zone disappear first, since the fermion states contributing to them are located on the periphery of the pockets. The susceptibility on the diagonals decreases more slowly, as the respective states are located near centers of the pockets. Though the size of the arcs in Fig.~\ref{Fig3} decreases with $\omega$, it remains finite up to zero frequency. Therefore in contrast to the superconducting phase the low-frequency susceptibility does not vanish in the pseudogap phase. Susceptibilities at the incommensurate maxima in the two phases as functions of $\omega$ are compared in Fig.~\ref{Fig5}.

The above-discussed difference in the behavior of the low-frequency susceptibility in the superconducting and pseudogap phases can account for the distinction between the underdoped and overdoped cuprates, if additionally a phase separation into superconducting and pseudogap regions is supposed in the former compounds. In contrast to the optimally and overdoped crystals, in underdoped cuprates below $T_c$ the superconducting gap in the susceptibility is not observed because the low-frequency magnetic response is determined by the pseudogap regions, while both regions contribute to the susceptibility for larger $\omega$. Thereby the incommensurability parameter -- the difference between the wave vector of the susceptibility maximum and {\bf Q} -- varies continuously from $\omega=0$ to higher frequencies and the dispersion of the maxima has the hourglass shape, as in optimally and slightly overdoped cuprates.

The above discussion allows us to propose the mechanism of the reorientation of the low-frequency susceptibility maxima from the axial to the diagonal direction, which is observed in LSCO for $x\lesssim 0.05$ \cite{Fujita11}. As mentioned previously, for large Fermi arc lengths the susceptibility is peaked at the momenta $(\pi,\pi\pm\kappa)$, $(\pi\pm\kappa,\pi)$ up to the smallest considered frequencies [see, e.g., Fig.~\ref{Fig2}(a)]. However, the Fermi arc length decreases nearly linearly with $x$ \cite{Yoshida}. As mentioned in the foregoing, the reduction of the arc length suppresses the susceptibility on the edge of the Brillouin zone more strongly than on the diagonal [see Fig.~\ref{Fig2}(b)]. For the used parameters $\chi''$ on the diagonal becomes larger than on the edge for $l\approx 0.6$. This value approximately corresponds to $x=0.04$ \cite{Yoshida}, which is close to the concentration of the maxima reorientation in LSCO. Notice that in the above calculations only the variation in the Fermi arc length was taken into account without regard for modifications in the hole band with $x$. The obtained reasonable value of the reorientation concentration apparently indicates that the Fermi arc length is the main parameter which determines this concentration. Let us recall that also for such small $x$ the dispersion of the susceptibility maxima has the hourglass shape.

\begin{figure}
\centerline{\resizebox{0.75\columnwidth}{!}{\includegraphics{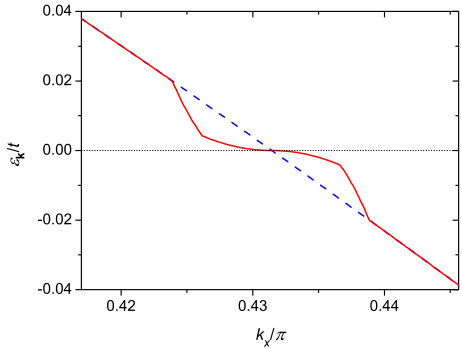}}}
\caption{The section of the hole dispersion along the diagonal of the Brillouin zone in the presence (the red solid line) and in the absence (the blue dashed line) of the cusp.} \label{Fig6}
\end{figure}

As seen in Fig.~\ref{Fig5}, with the used hole dispersion and $T=0$ the intensity of the incommensurate maximum decreases monotonously down to some finite value as $\omega$ is reduced. In some experiments in underdoped cuprates the growth of the susceptibility is observed with decreasing $\omega$. As an example of such a behavior, in Fig.~\ref{Fig5} the susceptibility measured in LSCO \cite{Kofu} is given. A possible reason for this frequency dependence may be a small cusp of the hole dispersion on the Fermi level. Such a cusp was obtained in the calculations of the hole dispersion of the 2D $t$-$J$ model for moderate $x$ \cite{Sherman97}. A part of the cusp -- a bending down of the dispersion -- is observed in some photoemission spectra of underdoped cuprates (see, e.g., Fig.~1 in Ref.~\cite{Terashima}). The cusp is shown in Fig.~\ref{Fig6}.
It increases the combined electron-hole density of states at low frequencies in the first term of Eq.~(\ref{impi}), which strengthens the susceptibility. The non-monotonic frequency dependence of the susceptibility caused by this small change of the hole dispersion is shown by filled triangles in Fig.~\ref{Fig5}. There is a clear similarity with the experimental dependence from Ref.~\cite{Kofu}. The deviation for $\omega\lesssim 1$~meV is presumably connected with the difference in temperatures in the experiment and calculations.

\begin{figure}
\centerline{\resizebox{0.75\columnwidth}{!}{\includegraphics{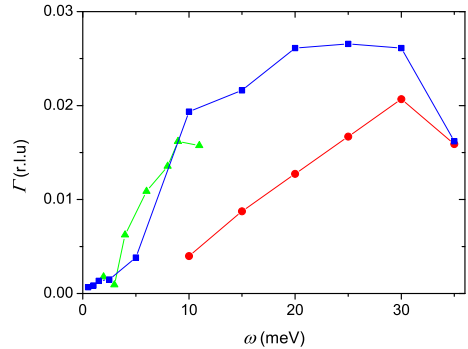}}}
\caption{The {\bf k} width $\Gamma$ of the incommensurate susceptibility maximum (half width at half maximum) as a function of $\omega$ in the superconducting (red circles) and pseudogap ($l=1.6$, blue squares) phases. Green triangles show the width measured in La$_{1.87}$Sr$_{0.13}$CuO$_4$ at $T=4$~K \protect\cite{Kofu}. The width is given in reciprocal lattice units.}\label{Fig7}
\end{figure}

In Ref.~\cite{Kofu} a rapid change of the {\bf k} width of the susceptibility maximum was revealed near the frequency $\omega=4$~meV in LSCO with $x\approx 0.13$ at $T=4$~K. Results of this work are shown in Fig.~\ref{Fig7} by green triangles. The widths obtained in our calculations for the superconducting and pseudogap phases are shown by red circles and blue squares, respectively. In the superconducting state the susceptibility maximum on the edge of the Brillouin zone disappears below $\omega=10$~meV. For larger frequencies the width grows nearly linearly at first and then, near $\omega_r\approx 0.08t=40$~meV, it starts to decrease, since the maxima transform gradually to a sharp resonance peak. In the pseudogap phase the number of states contributing to the susceptibility maximum is larger than in the superconducting state. As a consequence the widths of maxima in the former phase are considerably larger. As seen in Fig.~\ref{Fig7}, there is a steep rise of the width near $\omega=5$~meV. It is connected with the shoulder of the main maximum, which moves to the corner of the Brillouin zone [see Fig.~\ref{Fig2}(a)]. This shoulder is weak for small $\omega$, however it grows rapidly with frequency and determines the width for $\omega>5$~meV. This frequency dependence resembles the behavior of the linewidth observed in Ref.~\cite{Kofu}.

As seen in Fig.~\ref{Fig5}, for $T=0$ the susceptibility in the pseudogap phase tends to some finite value as $\omega\rightarrow 0$. From Eqs.~(\ref{impi}) and (\ref{imchi}) it follows that for finite temperatures $\chi''$ starts to decrease for $\omega_1\sim T$ and finally goes to zero at $\omega=0$. Usual temperatures of experiments of quasi-elastic neutron scattering are $2-4$~K (see, e.g., Refs.~\cite{Kofu,Haug}). Thus, $\omega_1\sim 10^{-1}$~meV, which is comparable to the energy resolution of these experiments. Nevertheless, the tendency for a decrease of $\chi''$ as the frequency becomes smaller than $T$ can be observed in some experimental spectra, for example, in the data reproduced in Fig.~\ref{Fig5}.

\section{Conclusion}
\label{conclusion}
Above we supposed that for temperatures $T<T_c$ the low-frequency magnetic response of underdoped cu\-p\-ra\-tes is determined by regions of the pseudogap phase in phase-separated crystals. It looks rather reasonable to extend this picture to the interpretation of the influence of modest magnetic fields and impurities on magnetic properties of these crystals below $T_c$. An external field, which is smaller than the upper critical field, enhances the quasi-elastic peaks at incommensurate positions in neutron scattering \cite{Haug,Katano,Lake}. In optimally doped LSCO such a field leads to a partial filling of the superconducting gap in the susceptibility \cite{Lake01}. These results suggest that the part of the crystal in vertex cores is in the pseudogap phase, and in the underdoped case the cores increase the volume occupied by this phase. The above supposition about the state of the vertex cores was already made earlier based on the analysis of tunneling experiments \cite{Renner,Levy} and model calculations \cite{Berthod}. Apparently manifestations of the pseudogap phase are observed not only in the vertex cores but also in their vicinity: the checkerboard pattern with periodicity of approximately four lattice spacings was observed inside and near the cores by scanning tunneling spectroscopy in slightly overdoped Bi$_2$Sr$_2$CaCu$_2$O$_x$ \cite{Hoffman,Matsuba}. A similar modulation was found above $T_c$, in the pseudogap phase, in Ref.~\cite{Vershinin}. As indicated in Ref.~\cite{Hoffman}, the electronic modulation is related to the magnetic modulation with twice the wavelength. In this picture, the nearly linear magnetic field dependence of the quasi-elastic neutron scattering at incommensurate momenta \cite{Haug} is related to the growing volume of the pseudogap regions, which do not overlap, while the tendency toward saturation \cite{Lake} arises with their overlapping. The observed strengthening of the quasi-elastic scattering by impurities \cite{Fujita} can be  connected also with regions of the pseudogap phase surrounding these impurities.

In summary, the magnetic susceptibility in the pseudogap phase of cuprate perovskites was calculated using the 2D $t$-$J$ model and the Mori projection operator technique. In contrast to the superconducting phase the susceptibility remains finite down to very low frequencies, which is the consequence of the Fermi arcs in the pseudogap phase. For frequencies immediately below the resonance frequency the magnetic response is incommensurate with maxima located on the edge of the Brillouin zone. For even lower frequencies the location of maxima depends on the length of the Fermi arcs: for long arcs the maxima are on the zone edge, while for short arcs they are on the diagonals. Together with the experimental fact that in LSCO the length of the Fermi arc is proportional to the hole concentration, this result explains the reorientation of the low-frequency susceptibility maxima with doping, observed in this crystal. The estimated value of the hole concentration, at which the reorientation occurs, is close to the experimental one. As in the superconducting state, the dispersion of the susceptibility maxima has an hourglass shape. Its lower part is connected with the maxima in the spin-excitation damping, while the upper part reflects the dispersion of spin excitations. The hourglass waist -- the resonance frequency -- corresponds to the gap in the spin-excitation dispersion due to the short-range antiferromagnetic order. A small cusp of the hole dispersion near the Fermi level, which is connected with the spin-polaron band, affords a growth of the incommensurate susceptibility maxima with decreasing frequency. Due to a peculiar shape of the susceptibility maximum and its variation with frequency in the pseudogap phase a sharp change of the {\bf k} width of the maximum occurs at some frequency. With the assumption of the phase separation into regions of superconducting and pseudogap phases in underdoped cuprates, these results provide explanation for the difference in the magnetic response of underdoped and overdoped lanthanum and yttrium cuprates. Below $T_c$ optimally and overdoped crystals are characterized by the superconducting gap in the magnetic susceptibility, while phase-separated underdoped samples have nonzero susceptibility down to the lowest measured frequencies at low temperatures. Moreover, in the latter crystals the susceptibility may grow with decreasing frequency. These peculiarities, as well as the observed rapid change of the {\bf k} width of the susceptibility maximum with frequency are reproduced in reasonable agreement with experiment, if we relate them to the pseudogap regions of the phase-separated crystal.

This work was supported by the project TLOFY0145 and by the European Regional Development Fund (project TK114).

\end{document}